\documentclass[twocolumn]{aastex63}

\usepackage{xcolor, ulem, soul}

\newcommand{\ngal}{$\mathrm{N}_\mathrm{gal} \ $}

\newcommand{\mgas}{$\mathrm{M}_\mathrm{gas} \ $}

\newcommand{\gsim}{\raisebox{-0.13cm}{~\shortstack{$>$ \\[-0.07cm]$\sim$}}~}

\accepted{3 March 2021}
\submitjournal{ApJL}

\shorttitle{A CO survey of SpARCS BCGs}
\shortauthors{Dunne et al.}
\graphicspath{{./}{figures/}}

\begin{document}

\title{A CO Survey of SpARCS Star-Forming Brightest Cluster Galaxies: Evidence for Uniformity in BCG Molecular Gas Processing Across Cosmic Time}

\author[0000-0002-5223-8315]{Delaney A. Dunne}
\affiliation{McGill Space Institute, Department of Physics, McGill University, 3600 rue University,  Montreal, QC H3A 2T8, Canada}

\author{Tracy M.A. Webb}
\affiliation{McGill Space Institute, Department of Physics, McGill University, 3600 rue University,  Montreal, QC H3A 2T8, Canada}

\author[0000-0003-1832-4137]{Allison Noble}
\affiliation{ASU School of Earth and Space Exploration, PO Box 871404, Tempe, AZ 85287, USA}

\author{Christopher Lidman}
\affiliation{Research School of Astronomy and Astrophysics, Australian National University, Canberra ACT 2611, Australia}
\affiliation{Centre for Gravitational Astrophysics, College of Science, The Australian National University, ACT 2601, Australia}

\author{Heath Shipley}
\affiliation{McGill Space Institute, Department of Physics, McGill University, 3600 rue University,  Montreal, QC H3A 2T8, Canada}

\author{Adam Muzzin}
\affiliation{Department of Physics and Astronomy, York University, 4700 Keele St., Toronto, ON MJ3 1P3, Canada}

\author{Gillian Wilson}
\affiliation{Department of Physics and Astronomy, University of California-Riverside, 900 University Avenue, Riverside, CA 92521, USA}

\author{H.K.C. Yee}
\affiliation{The David A. Dunlap Department of Astronomy and Astrophysics, University of Toronto, 50 St George St., Toronto, ON M5S 3H4, Canada}

\begin{abstract}

    We present ALMA CO (2-1) detections of 24 star-forming Brightest Cluster Galaxies (BCGs) over $0.2<z<1.2$, constituting the largest and most distant sample of molecular gas measurements in BCGs to date.  The BCGs are selected from the \textit{Spitzer} Adaptation of the Red-Sequence Cluster Survey (SpARCS) to be IR-bright and therefore star-forming. We find that molecular gas is common in star-forming BCGs, detecting CO at a detection rate of 80\% in our target sample of 30 objects. We additionally provide measurements of the star formation rate (SFR) and stellar mass, calculated from existing MIPS 24 $\mu$m and IRAC 3.6 $\mu$m fluxes, respectively. We find these galaxies have molecular gas masses of $0.7-11.0\times 10^{10}\ \mathrm{M}_\odot$, comparable to other BCGs in this redshift range, and specific star formation rates which trace the \citet{elbaz2011_ssfrMS} Main Sequence.
    We compare our BCGs to those of the lower-redshift, cooling-flow BCG sample assembled by \citet{edge2001detection} and find that at z $\lesssim 0.6$ the two samples show very similar correlations between their gas masses and specific SFRs. We suggest that, in this redshift regime, the $\sim10\%$ \citep{webb2015_SpARCS} of BCGs that are star-forming process any accreted molecular gas into stars through means that are agnostic to both their redshift and their cluster mass.

\end{abstract}


\section{Introduction} \label{sec:intro}
Galaxy clusters are extreme examples of hierarchical structure formation, forming the largest overdensities in the primordial universe. Brightest Cluster Galaxies (BCGs), as massive galaxies residing in the dense, violent centers of galaxy clusters, evolve through pathways that are a direct result of this environment.

In the latter stage of their evolution (after $\mathrm{z}\sim1$), the stellar mass of BCGs roughly doubles (\citealt{lidman2012evidence}). While this is expected to be mostly a dry process, done primarily through gasless mergers of the BCG with infalling cluster galaxies (\citealt{delucia_20017_SAMs}), some ($\sim$10\%, based on a 24 $\mu$m flux $\geq$ 100$\mu$Jy; \citealt{webb2015_SpARCS}) BCGs do show star formation at these redshifts. The mechanism driving this star formation is not clear. Possible scenarios include mergers, either major or minor, of the BCG with gas-rich infalling cluster galaxies, or large-scale `cooling flows' of gas from the intracluster medium (ICM) onto the BCG.

Galaxy clusters themselves are not a homogeneous population, complicating the search for a star formation mechanism in their centers. Early X-ray observations revealed that some cluster cores contain an ICM dense enough to cool hydrostatically on timescales much shorter than the Hubble time (e.g.~\citealt{fabian1977subsonic}). As the ICM cools and compresses, hot gas from outer regions of the cluster flows in to replace it, causing a cooling flow. Located at the center of the cluster, the BCG is often the recipient of gas from the cooling flow. Indeed, some star-forming BCGs seem to have star formation attributable to cooling flows. Examples include the Phoenix cluster at z $=0.596$ (\citealt{mcdonald2012_phoenixSFR}; \citealt{Russell_2017_phoenix}) and SpARCS104922.6+564032.5 (SpARCS1049; \citealt{Webb_2017_1049}; \citealt{hlavaceklarrondo2020_1049CC}) at z $=1.709$.

\begin{figure*}[ht!]
\centering
\includegraphics[width=0.9\linewidth]{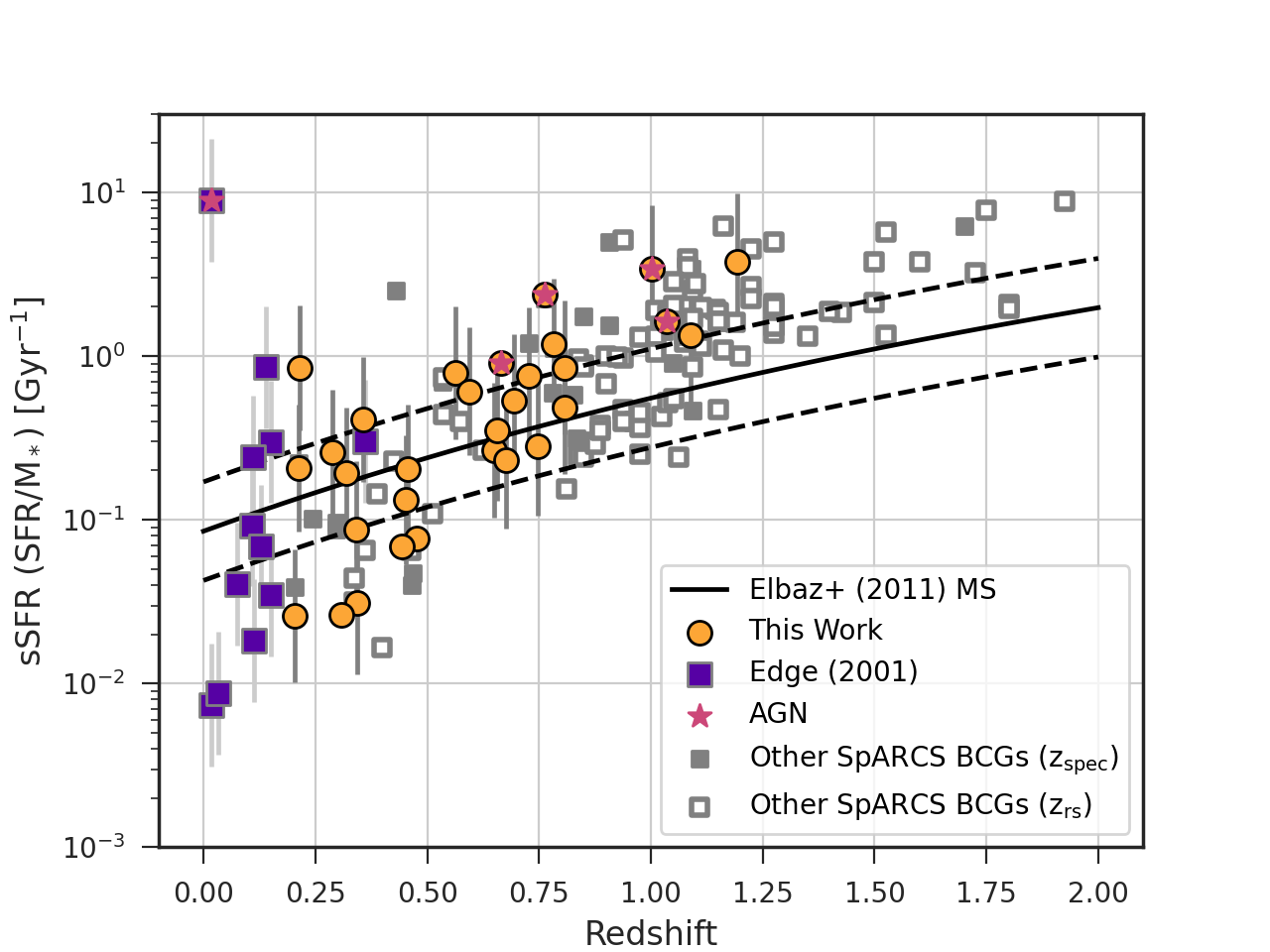}
\caption{The specific star formation rates (sSFRs) of the BCGs in our sample (orange circles), the \citet{edge2001detection} sample (purple squares) and the parent SpARCS BCG sample we draw from (grey squares; \citealt{webb2015_SpARCS}). Redshift values for this parent sample are principally the SpARCS RS-estimated redshifts (z$_\mathrm{RS}$; empty squares). Where available, we use spectroscopic redshifts (z$_\mathrm{spec}$; filled squares) from OzDES \citep{lidman2020_newozdes} or the literature (\citealt{menzel2016_sparcsz}; \citealt{weedmanHouck2009_sparcsz}; \citealt{swinbank2007_sparcsz}; \citealt{hernancaballero2009_sparcsz}; \citealt{SDSS_photometry_camera}; \citealt{sdssDR6}; \citealt{sdssDR7}). The \citet{elbaz2011_ssfrMS} sSFR MS is shown for comparison, with the scatter indicated with dashed lines. Galaxies are considered to be `starbursting' above the upper dashed line. Our sample of BCGs traces this \citet{elbaz2011_ssfrMS} MS, with scatter both above and below.  \label{fig:ssfrs}}
\end{figure*}

A key tool in deciphering star formation processes is cool molecular Hydrogen (H$_2$) gas, which acts as the fuel for future star formation in galaxies, and is commonly traced by the rotational transitions of CO \citep{solomon2005molecular}. However, few studies have systematically measured the molecular gas content in BCGs over a range of redshifts. We aim to constrain the processes driving BCG mass growth by studying molecular gas in a large sample of BCGs spanning 6 Gyr of cosmic time.

In this paper, we present new ALMA measurements of the CO (2-1) transition in 30 BCGs, selected from the SpARCS survey (described in \S\ref{ssec:parentSpARCS}). Our sample is large, containing 24 CO detections, and extends to much higher redshifts (z $\sim1.2$) than any previous statistical BCG survey. We introduce the SpARCS cluster sample, our ALMA observations, and our data reduction method in \S\ref{sec:observations}. In \S\ref{sec:results}, we contextualize our BCGs' molecular gas with other field and cluster measurements. \S\ref{sec:discussion} discusses the implications of these observations for BCG mass growth, and we present our conclusions in \S\ref{sec:conclusion}. We use standard cosmology ($\mathrm{H}_0 = 70$ km/s/Mpc; $\Omega_\mathrm{matter}=0.3$; $\Omega_\Lambda = 0.7$) throughout.

\section{Observations and Data Reduction}\label{sec:observations}

\subsection{The Parent SpARCS Sample}\label{ssec:parentSpARCS}

The BCGs studied here are drawn from clusters in the Spitzer Adaptation of the Red Sequence Cluster Survey (SpARCS; \citealt{muzzin2009spectroscopic}; \citealt{wilson2009spectroscopic}; \citealt{demarco2010spectroscopic}; \citealt{muzzin2012gemini}). SpARCS employs the same technique as the Red Sequence Cluster Survey (\citealt{gladders2005red}):  it identifies regions of over-densities in Red  Sequence (RS) galaxies by dual-filter imaging spanning the 4000\AA-break.  To do this SpARCS obtained wide-field $z^\prime$ imaging over the {\it Spitzer}-SWIRE Legacy Survey fields which, in combination with {\it Spitzer}-IRAC Channel 1 (3.6 $\mu$m) imaging, brackets the break above $z \gsim 1$. This simultaneously provides a reliable redshift estimate ($\delta z = (z_\mathrm{spec} - z_\mathrm{RS})/(1-z_\mathrm{spec})$ = 0.075, from $\sim$600 spectroscopic redshifts) through a fit to the location of the Red Sequence in color space. Note that though designed to be sensitive to $z> 1$ clusters,
the SpARCS catalog additionally contains many clusters at $z < 1$. Brightest Cluster Galaxies were identified at 3.6 $\mu$m, by taking the brightest galaxy within  $z - 3.6 \ \mu\mathrm{m}\pm0.5$ of the Red Sequence of the Cluster, and within 500 kpc of the centroid of the Red Sequence over-density.

\subsection{Target Selection}\label{ssec:targetselection}

We select as ALMA targets BCGs which a) lie in the southern/equatorial SWIRE fields ELAIS-S1 (ES1), CDFS, and XMM-LSS (XMM), b) have a confirmed spectroscopic redshift from OzDES (the Australian Dark Energy Survey), taken using the AAOmega spectrograph and 2dF fibre positioner on the 4m Anglo-Australian Telescope \citep{lidman2020_newozdes}, c) are IR-luminous ($>100\ \mu$Jy at $24 \ \mu$m) and therefore star-forming, d) come from SpARCS clusters with richness N$_\mathrm{gal} > $ 12\footnote{N$_\mathrm{gal}$ is defined as the background-subtracted number of galaxies (of any color) within 500 kpc of the cluster centre brighter than (M$^*_{3.6\mu\mathrm{m}} + 1$). }, or roughly M$_{200} > 1\times 10^{14}$M$_\odot$ (\citealt{wen2010erratum}; \citealt{wen2012catalog}; \citealt{capozzi2012kcorr}; \citealt{andreon2014_ngalmass}), and e) have CO (2-1) lines which fall into an ALMA frequency band. These leaves us with a sample of 33 BCGs.

\subsection{Observations}\label{ssec:observations}

\begin{figure*}[ht!]
\centering
\includegraphics[width=1.0\linewidth]{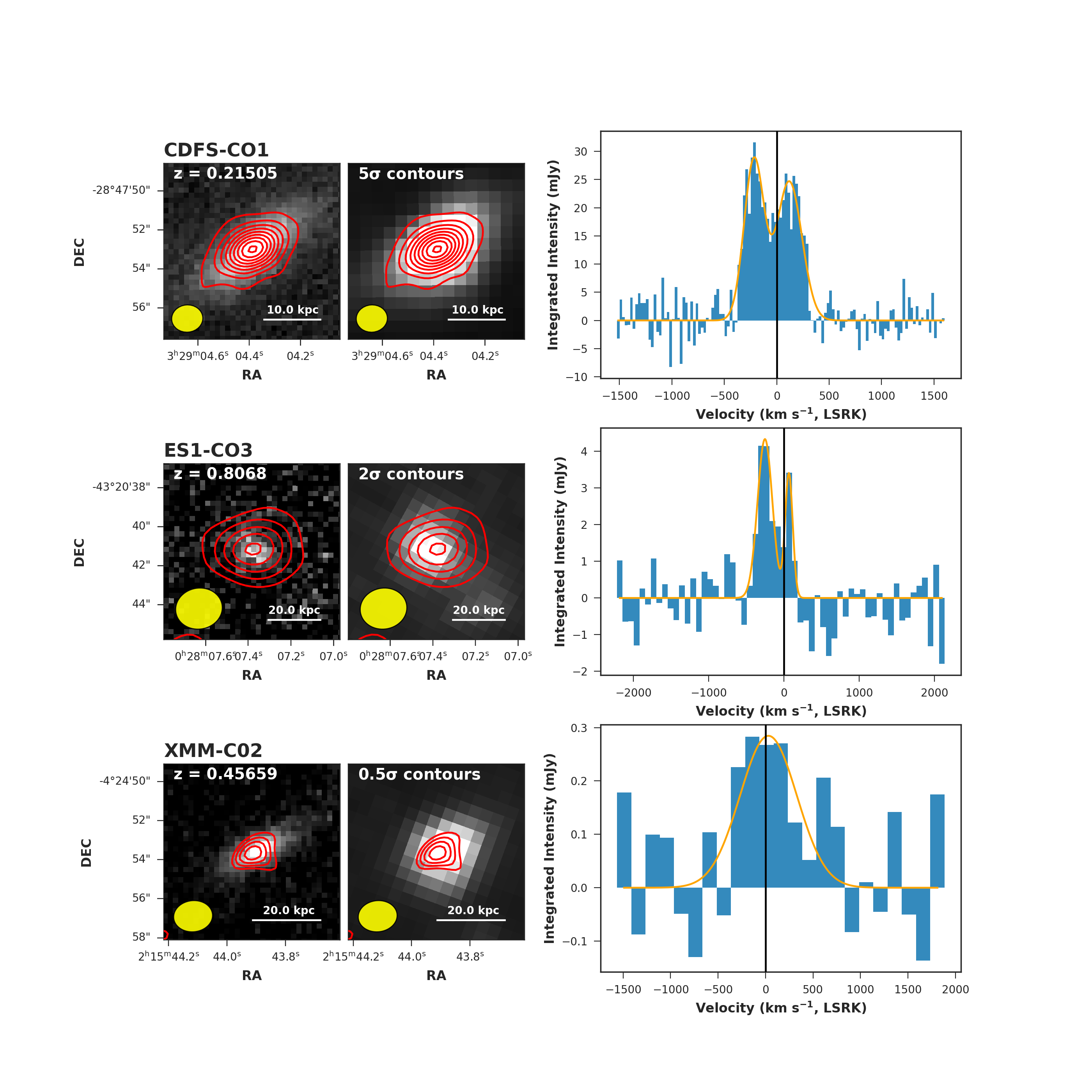}
\caption{High-S/N (top), medium-S/N (center), and low-S/N (bottom) examples of our galaxies detected in CO, spanning the redshift range of our sample. Postage stamps are 9$\arcsec$ $\times$ 9$\arcsec$, showing the ALMA CO(2-1) integrated-intensity maps in S/N contours (red) overlaid on DES z-band imaging (left) and IRAC Channel 1 (3.6 $\mu\mathrm{m}$) imaging (right). The ALMA synthesized beam is shown as the yellow ellipse. We show also the spectral profile for each galaxy, binned to the frequency used for calculations and centered around the OzDES spectroscpic redshift. One- or two-profile Gaussian fits are overlaid in orange. \label{fig:postagestamps}}
\end{figure*}

Our observations were performed on the Atacama Large Millimeter Array (ALMA; Program ID 2019.1.01529.S, P.I. Webb) between November 22 and December 15, 2019. 30 BCGs were observed over a total of $\sim15$ hours. Three BCGs were not observed because of the ALMA COVID-19 shutdown. We used ALMA bands 3, 4, and 5 (chosen to measure the CO (2-1) line) and configurations C43-1, C43-2, and C43-3. Data were calibrated using ALMA reduction pipeline scripts in CASA version 5.4.0. The nominal flux density calibration uncertainty for ALMA sources in these bands is $\sim 5 - 10\%$ \citep{remijan2019alma}. We cleaned the image cubes minimally with natural weighting and pixel sizes between 0\farcs2 and 0\farcs4, resulting in continuum-subtracted and primary beam-corrected images with 46\arcsec $\ $to 90\arcsec $\ $fields of view. We detected mostly unresolved CO emission at $\geq 3\sigma$ in 24 of 30 pointings, or at a detection rate of 80\%. We chose the spectral resolution of each image cube individually, based on the strength of the detection. We show examples of 3 BCGs from our sample, spanning the observed redshift range and the observed S/N range, in Figure \ref{fig:postagestamps}. A single BCG (CDFS-CO3) was additionally detected in radio continuum.

These observations were designed to be complete to a constant gas mass depth of \mgas $\geq 10^{10}$ M$_\odot$ at all redshifts. However, due to ALMA's minimum on-source integration time for a single science goal, many of the lower-redshift objects (z $\lesssim$ 0.6) were observed for longer than necessary to achieve this molecular gas mass limit, resulting in 3$\sigma$ image RMS depths between 1.2 mJy/beam (z $\sim 0.2$) and 0.6 mJy/beam (z $\sim 1.2$) over 100 km/s channels. The sensitivity to low gas masses thus increases at lower redshifts (see Figure \ref{fig:mgas_vs_redshift}).

\subsection{CO (2-1) Molecular Gas Measurements}\label{ssec:mgasmeasurements}

We determine galaxy gas masses in a three-step process. Firstly, we generate first-pass spectral profiles by placing an elliptical aperture with semi-major and semi-minor axes twice as large as those of the synthesized beam, and at the same orientation, at the center of each BCG. Spectra are fit with a single one-dimensional Gaussian profile. Secondly, we generate an integrated-intensity `Moment-0' map by collapsing the image cube over the velocity channels with significant emission as determined by this Gaussian fit to the spectral profile. We keep channels within 2$\sigma$ (2 standard deviations) of the mean. We then fit potential signal in the Moment-0 map to a two-dimensional, elliptical Gaussian profile. Finally, we create a second one-dimensional spectral profile, extracted from an elliptical aperture with the parameters of the two-dimensional fitted profile; the new axis sizes are now three times the Gaussian full width at half maximum (FWHM).

We determine the mass of gas present in each galaxy by fitting the second-pass spectral profile to either single- or double-peaked Gaussians, depending on which gives the better fit according to its reduced $\chi^2$ statistic. We take the area under this fit as the total line flux, and we convert this flux to a CO luminosity using \citet{solomon2005molecular} Equation 3. We use galactic CO-to-$\mathrm{H}_2$ conversion factors ($\alpha_{CO}=4.36\ M_\odot (\mathrm{K}\ \mathrm{km}\ \mathrm{s}^{-1}\mathrm{pc}^2)^{-1}$, $r_{21}=0.8$), as these are typical of galaxies on the star-forming Main Sequence (ex. \citealt{solomon1997molecular}; \citealt{freundlich2019phibss2}) and allow easy comparison with existing $\mathrm{H}_2$ masses in BCGs, such as those of \citet{edge2001detection} and \citet{castignani2020molecular}.
We note, however, that using different CO-to-$\mathrm{H}_2$ conversion factors (for example, $\alpha_\mathrm{CO} = 0.9$ and $\mathrm{r}_{21}=0.85$, typical of ultraluminous infrared galaxies; \citealt{bolatto2013co}) does not affect the results of this paper.

We treat the RMS noise in line-free channels ($\sim 7\sigma$ from the mean) of the spectral profile as the uncertainty in each flux value, and account for it while performing the least-squares Gaussian fits. We also search for CO emission in our 6 non-detections by stacking their spectra, centering each around the expected position of the CO line using the OzDES redshifts. The stack shows no significant emission, down to 3$\sigma$.

We determine gas mass limits on the 6 non-detections by collapsing the image cubes over a bandwidth equivalent to the average FWHM of the detected CO peaks ($356\ \mathrm{km}\ \mathrm{s}^{-1}$) and taking the $3\sigma$ RMS value as an upper limit to the flux.

\subsection{Stellar Masses and Star Formation Rates}\label{ssec:stellarmass_SFR}

We calculate the stellar mass of each galaxy using existing IRAC $3.6 \ \mu m$ photometry from SWIRE \citep{lonsdale2003swire} to determine the rest-frame \textit{K}-band luminosity. We use an SED template from \citet{bruzual2003stellar} to calculate the \textit{K}-correction, adopting an 11 Gyr simple stellar population with solar metallicity and a Salpeter Initial Mass Function (IMF; \citealt{salpeter_1955}). We note that adopting a template with a younger stellar population changes the average stellar mass values by at most 8\%. This is within uncertainty and does not affect our results significantly, as they are primarily comparative. We determine a mass-to-light ratio of 0.83 by comparing our sample of $3.6\ \mu m$-based values to more accurate values from a \textit{griz}/3.6$\mu$m/4.5$\mu$m (\citealt{Abbott_2018_DES}; \citealt{lonsdale2003swire}) SED fit using FAST \citep{kriek2018FAST}. We take the scatter in this relation (0.3 dex) as the uncertainty in the stellar masses.

We also calculate star formation rates (SFRs) from SWIRE-MIPS $24 \ \mu m$ flux values (\citealt{lonsdale2003swire}), using \citet{kennicutt1998_SFR} Equation 4 based on the luminosity at $24\  \micron$ ($L(24)$) from \citet{chary2001interpreting}. We note that this method is known to overestimate SFRs, particularly at high redshifts, but this generally occurs above the redshifts covered by our sample, at z $\sim 1.5$ \citep{elbaz2011_ssfrMS}. We achieve similar results with more modern SFR methods such as that of \citet{Rieke_2009}. We take the scatter of 0.15 dex described in \citet{kennicutt1998_SFR} to be the uncertainty in the SFRs. Because of SWIRE's 100 $\mu$Jy MIPS limit, our sample is not complete in SFR --- we can detect galaxies significantly ($\sim 1 \sigma$) below the \citet{elbaz2011_ssfrMS} MS only to z $\sim$ 0.6, and only starbursting galaxies above z $\sim$ 1.

We present the properties of our BCG sample, including M$_*$ and SFR values, in Table \ref{tab:results}.

\begin{figure*}[ht!]
\centering
\includegraphics[width=0.9\linewidth]{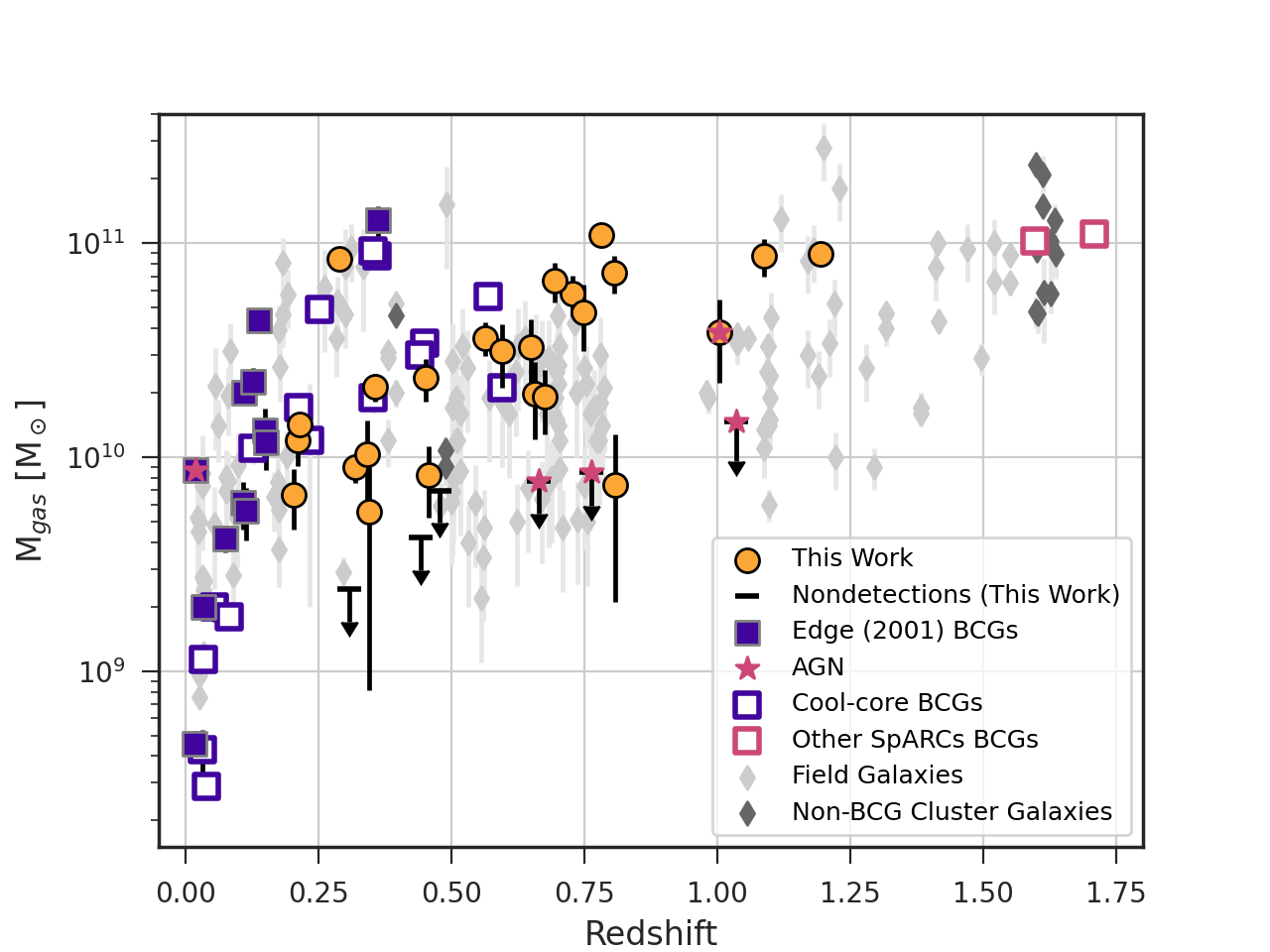}
\caption{The current state of molecular gas measurements in galaxies across the literature. Our BCG sample (orange circles) is compared to other BCG CO measurements (squares) and non-BCG cluster (dark grey diamonds) and field (light gray diamonds) galaxies taken from the literature (\S\ref{ssec:gasproperties}). In the past, CO observations of BCGs have overwhelmingly targeted BCGs in cool-core clusters (empty purple squares). Pink stars indicate the BCGs in our analysis which fall into the \citet{Donley_2012} AGN wedge. We pick out the \citet{edge2001detection} sample in particular (filled purple squares), as a large well-studied survey of low-redshift BCGs to compare to our data. The two pink squares are high-redshift BCGs in clusters whose dynamical state has not conclusively been determined --- SpARCS1049 (\citealt{Webb_2017_1049}) and J0225-303 (\citealt{noble2019_clustergalmgas}). Except for our own survey, we show only secure molecular gas detections, and not upper limits.  \label{fig:mgas_vs_redshift}}
\end{figure*}

\subsection{A Low-Redshift BCG Comparison Sample}\label{ssec:edge}

\citet{edge2001detection} introduces a sample of 34 BCGs at redshifts $\leq$ 0.45 in clusters selected from ROSAT to have cooling flows. Of these, 16 galaxies are detected in CO, through a combination of CO(3-2) measurements made using the A3i reciever on JCMT, and CO(1-0) measurements made using the IRAM 30m telescope. Additionally, 12 BCGs have available $3.6 \ \mu\mathrm{m}$ and $24\ \mu\mathrm{m}$ flux measurements. As the largest available sample of well-studied BCGs at low redshifts ($\mathrm{z}_\mathrm{mean} = 0.12 \pm 0.09$), belonging to clusters with known dynamical states, this is a convenient sample with which to compare our BCG detections.

We take the molecular gas masses from \citet{edge2001detection}, correcting the values for $H_0 = 70\ \mathrm{km} \ \mathrm{s}^{-1} \ \mathrm{Mpc}^{-1}$. Stellar masses and SFRs have been published for the Edge sample (\citealt{Rawle_2012}; \citealt{O_Dea_2008}); however, we choose to recalculate these values to allow for a homogeneous comparison between the Edge sample and our own. We therefore calculate both stellar masses and SFRs for these BCGs using the methods outlined in \S\ref{ssec:stellarmass_SFR}, including multiplying the SFR by the same FAST conversion factor, as the multiband photometry used in this SED fit is not available for the Edge galaxies. We take 3.6 $\mu$m flux values from \citet{quillen2008infrared}, \citet{Yamagishi_2010}, and \citet{landt_2010}), and MIPS 24 $\mu$m photometry from \citet{quillen2008infrared}, \citet{Yamagishi_2010}, \citet{shi_2005}, and the Spitzer Science Source List (\citealt{2010ASPC..434..437T}).

\subsection{AGN Contamination}\label{ssec:AGN}

To check for AGN contamination, we compare the IRAC colours of both our sample (from SWIRE; \citealt{lonsdale2003swire}) and the Edge sample (\citealt{quillen2008infrared}; \citealt{Yamagishi_2010}; \citealt{landt_2010}) to the criteria described in \citet{Donley_2012}. We find one CO-detected galaxy from each sample falls into the Donley AGN wedge, as well as the three nondetected galaxies at higher redshifts. These galaxies all show high sSFRs, and the $24 \ \mu m$ flux is likely dominated by the AGN; we thus exclude them from all further analysis but continue to show them on plots.

\section{Results}\label{sec:results}

The molecular gas properties of our BCG sample are presented in Table \ref{tab:results}. With the largest and highest-$z$ sample of significant molecular gas detections in BCGs assembled, we can investigate the reservoirs fueling their in-situ star formation.

\subsection{Gas properties compared to the larger galaxy population}
\label{ssec:gasproperties}

To place the gas in our BCGs in context, we compare in Figure \ref{fig:mgas_vs_redshift} the molecular gas reservoirs found in our sample to those of a range of other galaxy types\footnote{Non-BCG cluster galaxies are taken from \citet{noble2019_clustergalmgas}, \citet{noble_2017} and \citet{Jablonka_2013}. Field galaxies are taken from \citet{aravena2020_ASPECS}, \citet{Bauermeister_2013_EGNOG}, \citet{boogaard2020alma}, \citet{braun2011molecular}, \citet{brownson2020drives}, \citet{combes2013_polarringgalaxies}, \citet{Decarli_2016_ASPECS},  \citet{freundlich2019phibss2}, \citet{geach2011evolution}, and \citet{hatsukade2020_GRBhosts}. Other BCG detections are from \citealt{castignani2020molecular}, \citet{fogarty2019_MACS1931.8-2635bcg}, \citet{mcnamara2014_a1835bcg}, \citet{russell_2014_abell1664bcg}, \citet{Russell_2017_phoenix}, \citealt{salome_combes_2003}, \citet{tremblay2016_abell2597bcg}, and \citet{vantyghem2016_2A0335+096}.}.
We pick out the \citet{edge2001detection} sample described in \S\ref{ssec:edge} as our main comparison sample. We indicate also the two other BCGs selected from SpARCS with confident molecular gas detections --- SpARCS1049 (\citealt{Webb_2017_1049}), whose gas has recently been shown to be fed by a massive cooling flow (\citealt{hlavaceklarrondo2020_1049CC}) and J0225-303 (\citealt{noble2019_clustergalmgas}), which possibly represents a merging BCG based on rest-frame optical HST imaging, but whose cluster dynamical state has not been conclusively determined. The gap in coverage visible at $0.8\lesssim z \lesssim 1.0$ corresponds to the redshift range within which CO lines fall into the gap in the ALMA frequency coverage between bands 3 and 4.

Figure \ref{fig:mgas_vs_redshift} makes it immediately obvious that our sample reaches much higher redshift ranges than all previous BCG surveys. Few large-scale studies of molecular gas in BCGs have been undertaken, and those which do exist have been limited in distance by both CO instrument quality and the availability of cluster surveys at high redshifts. Above z $\sim 0.3$, most existing BCG gas masses come from targeted observations of known exceptional objects. These include the BCGs in Abell 1664 (z $=$ 0.128; \citealt{russell_2014_abell1664bcg}), Abell 1835 (z $=$ 0.252; \citealt{mcnamara2014_a1835bcg}), the Phoenix cluster (z $=$ 0.596; \citealt{Russell_2017_phoenix}), and SpARCS 1049 (z $=$ 1.709; \citealt{Webb_2017_1049}).  Previous BCG studies include \citet{edge2001detection} (described in detail in \S\ref{ssec:edge}), who detecteed 16 BCGs, \citet{salome_combes_2003}, who detected 6 BCGs, with some overlap with the \citet{edge2001detection} sample, and \citet{castignani2020molecular}, who detected 5 BCGs. Our sample of 24 BCGs is therefore large by comparison. Each of these studies had detection rates considerably lower than our own, likely because these studies either did not select their sample based on star formation (\citealt{edge2001detection}; \citealt{salome_combes_2003}), or used a much weaker SFR criterion (\citealt{castignani2020molecular}).

\begin{deluxetable*}{ccclhcchhhhhhcchcc}
\tablenum{1}
\tablecaption{Properties of the ALMA-observed BCGs. Spectroscopic redshifts (4) are taken from OzDES (c.f.~\S\ref{ssec:observations}). The offset of the CO (2-1) velocity from this OzDES measurement is given in column (5). Measurements of the velocity-integrated CO (2-1) flux and the resulting gas masses are described in \S\ref{ssec:mgasmeasurements}. The FWHM presented (8) are the single-Gaussian CO linewidths. SFR / M$_*$ measurements are described in \S\ref{ssec:stellarmass_SFR}. \label{tab:results}}
\tablewidth{0pt}
\tablehead{
\colhead{SWIRE} & \colhead{RA} & \colhead{DEC} & \colhead{$z_\mathrm{OzDES}$} & \nocolhead{$z_\mathrm{CO}$} & \colhead{$v_\mathrm{CO}$ Offset} & \colhead{$S_\mathrm{CO}\Delta v$} & \nocolhead{Integration} & \nocolhead{Pixel} & \nocolhead{FoV} & \nocolhead{RMS} & \nocolhead{Channel} & \nocolhead{L$_{CO(2-1)}$} & \colhead{M$_\mathrm{gas}$} & \colhead{FWHM} & \nocolhead{Best-fit} & \colhead{SFR} &
\colhead{$\mathrm{M}_*$}\\
\colhead{Patch} & \colhead{[deg]} & \colhead{[deg]} & \colhead{} & \nocolhead{} & \colhead{[km s$^{-1}$]} & \colhead{[Jy km s$^{-1}$]} &
\nocolhead{Time (s)} & \nocolhead{Scale ["]} & \nocolhead{["]} & \nocolhead{[mJy $\mathrm{beam}^{-1}$]} & \nocolhead{Width [km $\mathrm{s}^{-1}$]} & \nocolhead{[$\times 10^9, \ \mathrm{L}_\odot$]} & \colhead{[$\times 10^{10}, _{ \ } \mathrm{M}_\odot$]} & \colhead{[km s$^{-1}$]} & \nocolhead{model} & \colhead{[$\mathrm{M}_\odot \ \mathrm{yr}^{-1}$]} &
\colhead{[$\times 10^{10}, _{ \ }\mathrm{M}_\odot$]}
}
\decimalcolnumbers
\startdata
XMM-CO1 & 33.65927 & -3.62736 & 1.0033 & $1.0058 \pm 0.0007$ & $187 \pm 100$ & $0.50 \pm 0.15$ & 1699.8 & 0.3 & 75.0 & 0.433 & 100 & $5.9\pm 2.5$ & $3.8 \pm 1.6$ & $147 \pm 84 \ $ & 1 & $670^{+280}_{-200}$ & $22.9^{+14.0}_{-8.7}$ \\
XMM-CO2 & 33.93287 & -4.41511 & 0.4566 & $0.4556 \pm 0.0003$ & -$150 \pm 150 \ $ & $0.51 \pm 0.20$ & 303.0 & 0.2 & 56.0 & 0.384 & 150 & $1.3 \pm 0.5$ & $0.8 \pm 0.3$ & $531 \pm 127$ & 1 & $27^{+11}_{-8}$ &  $14.8^{+9.1 \ }_{-5.6}$  \\
XMM-CO3 & 34.64331 & -5.01232 & 0.6410 & $0.6513 \pm 0.0003$ & $162 \pm 75_{\ }$ & $1.09 \pm 0.26$ & 301.8 & 0.3 &  63.6 & 0.398 &  75 & $5.0 \pm 1.7$ & $3.3 \pm 1.1$ & $133 \pm 63 \ $ & 1 & $34^{+14}_{-10}$ & $14.5^{+8.8 \ }_{-5.5}$ \\
XMM-CO4 & 34.79386 & -3.72621 & 0.7634 & - & - & $<0.26$ & 516.0 & 0.3 & 67.2 & 0.779 & 300 & $< 0.5$ & $<0.9$ & - & - & $410^{+169}_{-120}$ & $17.4^{+8.7 \ }_{-5.8}$ \\
XMM-CO5 & 34.87044 & -4.11663 & 0.7826 & $0.7824 \pm 0.0001$ & -$17 \pm 25 \ $ & $2.36 \pm 0.30$ &  303.0 & 0.2 & 50.0 & 0.663 & 100 & $1.0 \pm 0.3$ & $11.0 \pm 1.3 \ $ & $271 \pm 21 \ $ & 2 & $140^{+60 \ }_{-40}$ & $13.4^{+8.2 \ }_{-5.1}$ \\
XMM-CO6 & 34.92392 & -4.00916 & 0.2043 & $0.2042 \pm 0.0001$ & -$27 \pm 100$ & $1.95 \pm 0.57$ & 303.0 & 0.2 & 50.0 & 0.663 & 100 & $1.0 \pm 0.3$ & $0.7 \pm 0.3$ & $473 \pm 84 \ $ & 1 & $3.0^{+1.2}_{-0.9}$ & $12.4^{+7.5 \ }_{-4.7}$ \\
XMM-CO7 & 35.22837 & -3.54931 & 1.0346 & - & - & $<0.28$ & 2001.0 & 0.4 & 86.4 & 0.248 & 300 & $< 0.9$ & $< 1.7$ & - & 1 & $420^{+168}_{-120}$ & $26.0^{+13.0}_{-8.6}$ \\
XMM-CO8 & 36.11233 & -5.60891 & 0.2120 & $0.2112 \pm 0.0001$ & -$164 \pm 50 \ \ $ & $3.88 \pm 0.67$ & 301.8 & 0.2 & 46.4 & 0.535 & 25 & $1.8 \pm 0.4$ & $1.2 \pm 0.3$ & $503 \pm 42 \ $ & 2 & $11^{+4 \ }_{-3}$ & $6.5^{+4.0}_{-2.5}$ \\
\hline
ES1-CO1 & 10.26937 & -44.50626 & 0.3446 & $0.3477 \pm 0.0003$ & $517 \pm 200$ & $0.65 \pm 0.55$ & 301.8 & 0.2 & 52.0 & 1.200 & 200 & $0.8 \pm 0.7$ & $0.6 \pm 0.5$ & $188 \pm 169$ & 1 & $6.1^{+2.5}_{-1.8}$ & $22.2^{+13.5}_{-8.4}$ \\
ES1-CO2 & 6.97816 & -43.27981 & 1.0881 & $1.0866 \pm 0.0006$ & -$103 \pm 75_{\ \ \ }$ & $0.97 \pm 0.20$ & 1483.8 & 0.3 & 81.0 & 0.193 & 75 & $10.3 \pm 2.2 \ $ & $8.7 \pm 1.7$ & $191 \pm 63 \ $ & 2 & $75^{+31}_{-22}$ & $6.6^{+4.1}_{-2.5}$ \\
ES1-CO3 & 7.03042 & -43.3448 & 0.8068 & $0.8059 \pm 0.0001$ & -$84 \pm 25 \ $ & $1.48 \pm 0.29$ & 576.0 & 0.3 & 75.0 & 0.330 & 75 & $11.2 \pm 2.2 \ $ & $7.3 \pm 1.4$ & $403 \pm 21 \ $ & 2 & $73^{+30}_{-21}$ & $9.8^{+6.0}_{-3.7}$ \\
ES1-CO4 & 8.26314 & -42.72425 & 0.7281 & $0.7296 \pm 0.0004$ & $147 \pm 100$ & $1.37 \pm 0.25$ & 516.0 & 0.3 & 67.2 & 0.334 & 100 & $8.9 \pm 1.9$ & $5.8 \pm 1.3$ & $541 \pm 84 \ $ & 1 & $42^{+17}_{-12}$ & $6.5^{+4.0}_{-2.5}$ \\
ES1-CO5 & 8.70355 & -42.23410 & 0.4776 & - & - & $<0.41$ & 301.8 & 0.25 & 62.5 & 1.074 & 300 & $< 0.4$ & $<0.7$ & - & - & $12^{+6 \ }_{-4}$ & $15.6^{+7.8 \ }_{-5.2}$ \\
ES1-CO7 & 8.81504 & -42.78926 & 0.6567 & $0.6538 \pm 0.0007$ & -$314 \pm 200 \ $ & $0.59 \pm 0.21$ & 301.8 & 0.3 & 75.0 & 0.400 & 200 & $3.1 \pm 1.2$ & $2.0 \pm 0.8$ & $394 \pm 169$ & 1 & $28^{+11}_{-8}$ & $9.4^{+5.7}_{-3.6}$ \\
ES1-CO8 & 8.81970 & -42.68564 & 0.5640 & $0.5630 \pm 0.0001$ & -$113 \pm 50 \ \ $ & $1.58 \pm 0.34$ & 301.8 & 0.3 & 75.0 & 0.430 & 50 & $5.6 \pm 1.0$ & $3.6 \pm 0.6$ & $253 \pm 42 \ $ & 2 & $38^{+15}_{-11}$ & $5.5^{+3.4}_{-2.1}$ \\
ES1-CO9 & 9.41918 & -43.65286 & 1.1937 & $1.1924 \pm 0.0002$ & -$83 \pm 25 \ $ & $0.83 \pm 0.11$ & 1272.0 & 0.3 & 90.0 & 0.225 & 25 & $13.8 \pm 1.7 \ $ & $8.9 \pm 1.1$ & $156 \pm 21 \ $ & 1 & $250^{+100}_{-70}$ & $7.6^{+4.8}_{-2.9}$ \\
ES1-CO10 & 9.76238 & -44.36451 & 0.7474 & $0.7472 \pm 0.0004$ & $ \ $-$23 \pm 100$ & $1.13 \pm 0.39$ & 516.0 & 0.3 & 75.0 & 0.326 & 100 & $7.3 \pm 2.5$ & $4.8 \pm 1.6$ & $494 \pm 84 \ $ & 1 & $37^{+15}_{-10}$ & $15.2^{+9.3 \ }_{-5.8}$ \\
\hline
CDFS-CO1 & 52.26829 & -28.79808 & 0.2891 & $0.2888 \pm 0.0001$ & -$50 \pm 25 \ $ & $14.33 \pm 1.23$ & 303.0 & 0.2 & 50.0 & 0.620 & 50 & $13.0 \pm 1.0 \ $ & $8.4 \pm 0.7$ & $558 \pm 21 \ $ & 2 & $35^{+14}_{-10}$ & $15.8^{+9.6 \ }_{-6.0}$ \\
CDFS-CO2 & 52.90532 & -28.40428 & 0.2151 & $0.2159 \pm 0.0001$ & $170 \pm 25 \ $ & $4.41 \pm 0.27$ & 303.0 & 0.2 & 50.0 & 0.530 & 50 & $2.2 \pm 0.1$ & $1.42 \pm 0.09$ & $144 \pm 21 \ $ & 1 & $35^{+14}_{-10}$ & $4.9^{+3.0}_{-1.8}$ \\
CDFS-CO3 & 53.31982 & -26.89622 & 0.8076 & $0.810 \pm 0.001$ & $237 \pm 200$ & $0.18 \pm 0.39$ & 547.2 & 0.3 & 75.0 & 0.340 & 175 & $1.1 \pm 0.8$ & $0.7 \pm 0.5$ & $564 \pm 169$ & 1 & $54^{+22}_{-15}$ & $12.9^{+7.9 \ }_{-4.9}$ \\
CDFS-CO4 & 53.46281 & -27.33505 & 0.4511 & $0.4508 \pm 0.0002$ & -$53 \pm 75 \ $ & $1.59 \pm 0.49$ & 303.0 & 0.2 & 56.0 & 0.957 & 75 & $3.6 \pm 0.8$ & $2.4 \pm 0.5$ & $493 \pm 63 \ $ & 2 & $24^{+10}_{-7}$ & $20.6^{+12.6}_{-7.8}$ \\
CDFS-CO5 & 53.48398 & -27.25938 & 0.6948 & $0.6947 \pm 0.0004$ & -$13 \pm 100$ & $1.87 \pm 0.50$ & 303.0 & 0.3 & 75.0 & 0.360 & 100 & $10.3 \pm 2.2 \ $ & $6.7 \pm 1.4$ & $432 \pm 84 \ $ & 2  & $54^{+22}_{-15}$ & $11.5^{+7.1 \ }_{-4.4}$ \\
CDFS-CO6 & 53.55353 & -28.41692 & 0.6643 & - & - & $<0.23$ & 303.0 & 0.3 & 67.2 & 0.350 & 300 & $< 0.4$ & $<0.8$ & - & 1 & $180^{+69 \ }_{-50}$ & $20.1^{+10.0}_{-6.7}$ \\
CDFS-CO7 & 53.72804 & -27.09096 & 0.6760 & $0.6770 \pm 0.0008$ & $105 \pm 200$ & $0.55 \pm 0.18$ & 301.8 & 0.3 & 67.2 & 0.320 & 200 & $2.9 \pm 1.0$ & $1.9 \pm 0.6$ & $448 \pm 169 \ $ & 1 & $23^{+9 \ }_{-6}$ & $11.6^{+7.1 \ }_{-4.4}$ \\
CDFS-CO8 & 53.77638 & -29.26097 & 0.5950 & $0.5966 \pm 0.0006$ & $185 \pm 200$ & $1.18 \pm 0.39$ & 301.8 & 0.3 & 67.2 & 0.374 & 200 & $4.8\pm 1.6$ & $3.1 \pm 1.0$ & $471 \pm 169 \ $ & 1 & $73^{+30}_{-21}$ & $13.7^{+8.3 \ }_{-5.2}$ \\
CDFS-CO9 & 53.79709 & -27.77951 & 0.4428 & - & - & $<0.15$ & 301.8 & 0.2 & 56.0 & 0.366 & 300 & $< 0.2$ & $<0.4$ & - & - & $11^{+4 \ }_{-3}$ & $16.0^{+8.0 \ }_{-5.3}$ \\
CDFS-CO10 & 53.83965 & -29.61118 & 0.3079 & - & - & $<0.36$ & 301.8 & 0.2 & 54.0 & 0.511 & 300 & $< 0.2$ & $<0.3$ & - & - & $3.9^{+1.5}_{-1.1}$ & $14.8^{+7.4 \ }_{-4.9}$ \\
CDFS-CO11 & 53.86995 & -27.28562 & 0.3189 & $0.3190 \pm 0.0001$ & $ \ 9 \pm 25$ & $1.23 \pm 0.19$ & 303.0 & 0.2 & 54.0 & 0.430 & 25 & $1.4 \pm 0.3$ & $0.9 \pm 0.1$ & $56 \pm 21$ & 1 & $9.5^{+3.9}_{-2.8}$ & $5.6^{+3.4}_{-2.1}$ \\
CDFS-CO12 & 53.89642 & -27.26368 & 0.3422 & $0.3427 \pm 0.0003$ & $ \ 73 \pm 200$ & $0.90 \pm 0.46$ & 301.8 & 0.2 & 54.0 & 0.431 & 200 & $1.6 \pm 0.7$ & $1.0 \pm 0.5$ & $447 \pm 169$ & 1  & $6.1^{+2.5}_{-1.8}$ & $7.9^{+4.8}_{-3.0}$ \\
CDFS-CO13 & 54.07105 & -28.38037 & 0.3567 & $0.3562 \pm 0.0001$ & -$93 \pm 25 \ $ & $2.31 \pm 0.34$ & 301.8 & 0.2 & 54.0 & 0.370 & 25 & $3.3 \pm 0.5$ & $2.1 \pm 0.3$ & $327 \pm 21 \ $ & 2 & $36^{+15}_{-10}$ & $10.3^{+6.2 \ }_{-3.9}$ \\
\enddata
\end{deluxetable*}

\subsection{Direct comparison with lower-redshift BCGs}
\label{ssec:edgecomparison}

\begin{figure*}[ht!]
\centering
\includegraphics[width=\linewidth]{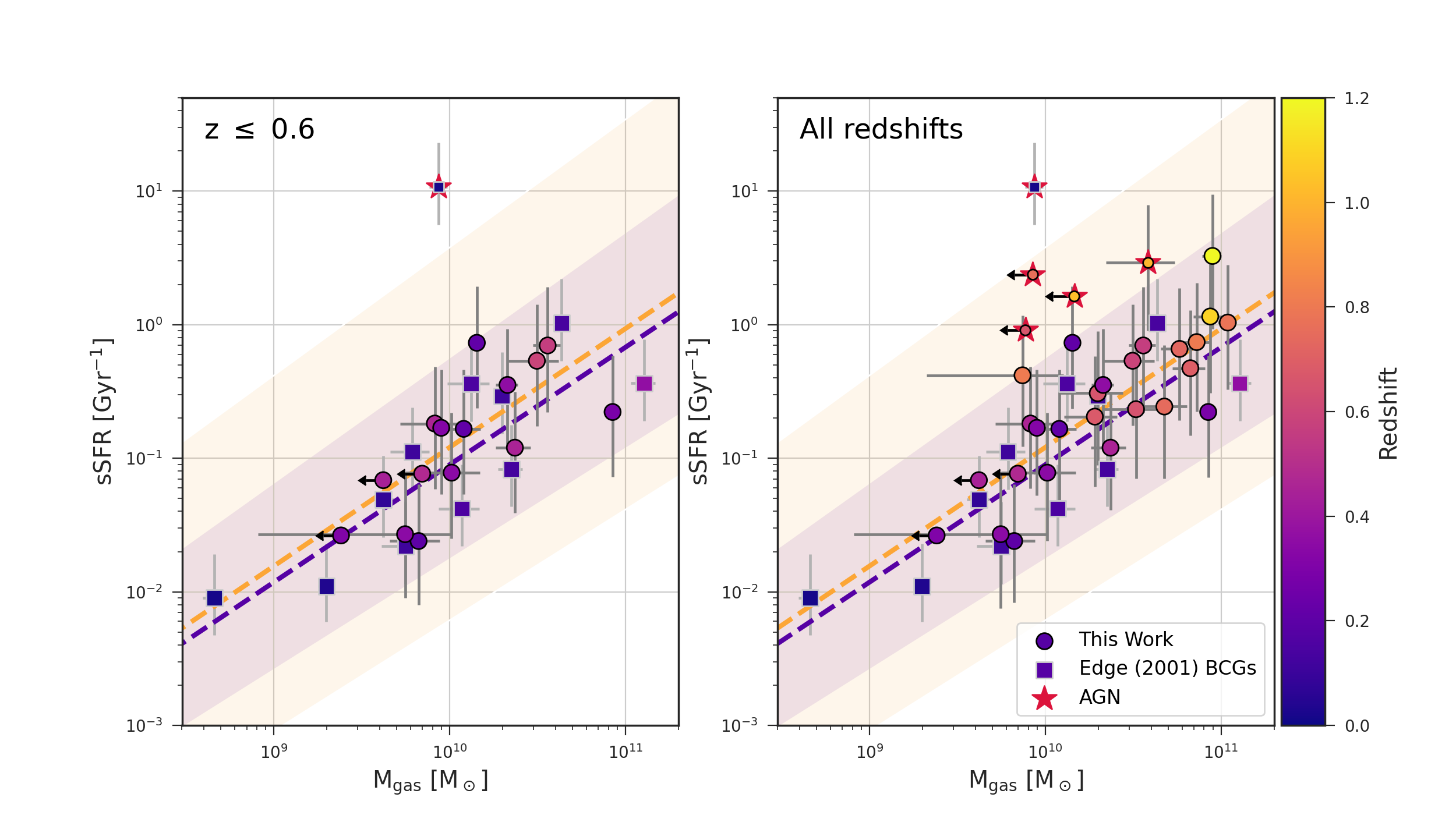}
\caption{A comparison of the properties of BCGs found in this study (circles) to those reported in \citet{edge2001detection} (squares). Points in both panels are colored according to their redshift. We show sSFR plotted against gas mass for both samples. The left panel contains only galaxies at z $\leq$ 0.6 to prevent selection effects in either gas mass or SFR. ODR fits to the BCGs introduced in this work (orange) and to the \citet{edge2001detection} BCGs (purple) are shown with their $1\sigma$ confidence regions shaded. Fits to BCGs at z $\leq$ 0.6 are also shown in the right panel, with the entire sample overlaid for reference. Galaxies not detected in CO (arrows) and those known to contain AGN (stars) are shown for reference, but are not included in the fits.
\label{fig:mgas_vs_ssfr}}
\end{figure*}

In the field, tight correlations are known to exist between a galaxy's SFR, M$_*$, and gas mass (e.g. \citealt{genzel2010_Scalingrelations}), suggesting a uniformity in baryon processing by field galaxies. To test if similar correlations exist in BCGs, we compare the sSFRs and molecular gas masses of our sample to those BCGs detected by  \citet{edge2001detection} in Figure \ref{fig:mgas_vs_ssfr}. Due to the sSFR depth limitations discussed in \S\ref{ssec:stellarmass_SFR}, we first analyze only the BCGs in our sample with redshifts below 0.6 (left panel). To determine if the two samples exhibit the same sSFR/\mgas properties, we fit a straight line to each sample separately in logarithmic space, using Orthogonal Distance Regression (ODR) to account for uncertainties existing in both axes. We show fits, and their 1$\sigma$ confidence regions, in Figure \ref{fig:mgas_vs_ssfr}. With (logarithmic) slopes of $0.9 \pm 0.3$ and $0.9 \pm 0.2$ and offsets of $-10 \pm 3$ and $-10 \pm 2$, respectively, our sample of BCGs and the \citet{edge2001detection} sample display the same correlation below z $=0.6$. As the slope in this plot is log(SFR/M$_*$) / log(M$_\mathrm{gas}$), a quantity related to the star formation efficiency (SFR/M$_\mathrm{gas}$), this suggests the two samples are processing molecular gas through similar means.

In the right panel of Figure \ref{fig:mgas_vs_ssfr}, we plot the complete sample over all redshifts. Much of our sample to z $\sim$ 1.0 falls into the $1\sigma$ confidence region of the fit to the \citet{edge2001detection} BCGs, suggesting that this similarity in gas processing could persist to higher redshifts, but this is inconclusive due to the small sample size and potential selection effects from limited sensitivity. The BCGs at redshifts above z $\sim$ 1.0 seem to fall above this relation, suggesting a possible change in gas processing, but this is again somewhat speculative due to the limits of the sample. In both panels the CO nondetections are included for comparison. We note the three nondetections which fall off the plotted correlation all have sSFRs which are likely inflated by the presence of AGN.

While we have worked to control for all potential systematic sSFR/\mgas differences between the two samples, there do remain several limitations. These include the different CO line transitions used in the \mgas values calculated by \citet{edge2001detection}, the difficulty defining selection effects in this sample (which includes measurements from several different sources), and the incompleteness of our own sample.  Additionally, while several large samples or field galaxies with molecular gas measurements do exist, such as ASPECS \citep{Decarli_2016_ASPECS}, PHIBSS1 \citep{tacconi2013phibss}, and PHIBBS2 \citep{freundlich2019phibss2}, few of these galaxies have both the 3.6 $\mu$m and 24 $\mu$m measurements necessary for rigorous comparison with the BCGs in this work, and those that do tend to fall above the MS. Ultimately, more CO data is required, for BCGs and field galaxies, particularly at high redshifts.

\section{Discussion}\label{sec:discussion}

We measure CO (2-1) in a sample of 30 star-forming BCGs, chosen from SpARCS. We find that 80\% of our BCGs are detected in CO, suggesting that molecular gas is common in star-forming BCGs at high redshifts. While this is seemingly in contrast with the results of similar BCG surveys such as \citet{castignani2020molecular}, we note that these previous studies imposed no SFR limit on their samples (see \S\ref{ssec:gasproperties}). We specifically target star-forming galaxies, and new stars cannot form without fuel.

Semi-Analytic Models suggest that strangulation, ram-pressure stripping, and other cluster effects actively remove molecular gas from galaxies in dense environments (e.g. \citealt{delucia_20017_SAMs}; \citealt{gavazzi2006_clustergasdef}), and therefore limit star formation. It is thus necessary to account for two different processes when analyzing BCG star formation --- the process by which gas is deposited onto the BCG, and the mechanism by which the BCG converts that gas into stars. This distinction between gas feeding and processing in BCGs may have important implications, especially as similar theories have been proposed for the general galaxy population (c.f.~\citealt{behroozi2019_halomass}).

In Figure \ref{fig:mgas_vs_ssfr}, we show that our BCG sample at z $\leq 0.6$ has a remarkably similar correlation between M$_\mathrm{gas}$ and sSFR to that of \citet{edge2001detection}, with nearly identical slopes. This implies that the two BCG samples are both processing their in-situ molecular gas into stars with the same efficiency for a given stellar mass. This correlation exists at z $\leq 0.6$, and possibly to higher redshifts (z $\sim1.0$), although this is uncertain due to the limiting depth of the sample, suggesting that many star-forming BCGs at $0<z\leq0.6$ may regulate the formation of these stars from gas in the same way, independent of their redshift or cluster properties. This is a picture of galaxy star formation which has been proposed for non-BCGs (e.g.~\citealt{bouche2010gasregulator}; \citealt{dekel2014analytic}).

Similarities in star-formation processes between the \citet{edge2001detection} sample and our own have the potential to be very interesting, as the two samples differ in both selection method and cluster mass, in addition to the redshift difference discussed in \S\ref{ssec:edge}. The \citet{edge2001detection} sample contains only BCGs from known cool-core clusters. In fact, all previous large BCG studies have selected galaxies either explicitly from known cool-core clusters (\citealt{edge2001detection}; \citealt{salome_combes_2003}), or from X-ray surveys (\citealt{castignani2020molecular}). The parent, red-sequence selected SpARCS sample does not select clusters based on ICM state, and should therefore exhibit no cool-core selection effects. While the cool-core fraction varies depending on the selection criteria used, it should be $\lesssim 0.3$ in the redshift range we study \citep{McDonald_2013CCfraction}.

We do note, however, that our BCG sample is selected on star formation, which may bias it towards cool-core clusters (e.g. \citealt{Fogarty_2015_CLASHsf}). If our galaxies are all cool-core cluster BCGs, and therefore fed by cooling flows, this is an interesting result on its own, as additional confirmation that the star formation we see in BCGs at z $\lesssim 0.6$ may be the result of those BCGs being fed molecular gas principally by cooling flows, instead of other possible scenarios such as gas-rich mergers. Unfortunately, we lack the X-ray measurements necessary to confirm the dynamical state of our clusters.

These speculations should not affect our main result --- \mgas is a measurement of the amount of gas already in the galaxy. However, the CO-to-H$_2$ conversion factor used to determine molecular gas mass depends, among other things, on gas density (e.g.~\citealt{bolatto2013co}). If some of our BCGs are dynamically unrelaxed post-mergers, this conversion factor could be affected, changing the relation. It is possible that we see this effect in Figure \ref{fig:mgas_vs_ssfr} at high redshifts, where BCGs tend to appear above the relation. This is in line with past work such as \citet{mcdonald2016star}, which suggests that the mechanism fueling star formation changes from primarily cooling flows to gas-rich mergers in BCGs at z $\gsim 0.6$.

We cannot compare the cluster halo masses of our sample to those of \citet{edge2001detection} in depth, as we have no equivalent measurements to compare. We do, however, have cluster richnesses ($\mathrm{N}_\mathrm{gal}$) measured from SpARCS (\citealt{webb2015_SpARCS}) for our sample, and halo masses based on SDSS DR7 (\citealt{popesso2020_edgemcluster}; \citealt{yang2005galaxy}) for the \citet{edge2001detection} BCGs. As \ngal $=12$ is roughly equal to M$_{200} \sim 1\times10^{14}$ (e.g. \citealt{andreon2014_ngalmass}), our clusters have an average cluster mass $\sim (1.5 \pm 0.7)\times10^{14} \ \mathrm{M}_\odot$, compared to an average cluster mass of $\sim (1.3 \pm 0.9)\times 10^{15}\ \mathrm{M}_\odot$ for the \citet{edge2001detection} sample. The clusters presented in \citet{edge2001detection} are therefore almost an order of magnitude more massive than ours.

It has been suggested that halo mass is directly related to central galaxy star formation, with the fraction of quenched central galaxies increasing with host halo mass (e.g.~\citealt{behroozi2019_halomass}). However, the star formation efficiency for galaxies at a given distance from the MS has been shown not to change with host halo mass \citep{popesso2020_edgemcluster}. We select only star-forming galaxies (i.e., galaxies at a fixed distance from the MS; Figure \ref{fig:ssfrs}), so our values are unaffected by the fraction of quenched central galaxies at a given halo mass, and it is therefore unsurprising that the two BCG samples line up well despite their differing cluster masses.

\section{Conclusions}\label{sec:conclusion}
We present the largest and highest-redshift study of molecular gas in BCGs to date. Using ALMA bands 3, 4, and 5, we study CO (2-1) in 30 SpARCS BCGs. We summarize our results as follows:
\begin{enumerate}
    \item We find that molecular gas is common in star-forming BCGs at $0.2\leq \mathrm{z}\leq 1.2$, detecting gas reservoirs in 80\% of our sample of 30 galaxies.

    \item We note a clear M$_\mathrm{gas}$-sSFR trend is apparent in our sample. Below z $=0.6$ this trend is consistent to $1\sigma$ with the sample of BCGs presented in \citet{edge2001detection}. We therefore propose a scenario by which most BCGs below z $\sim0.6$ process molecular gas into stars at a rate dependent on stellar mass, through mechanisms independent of redshift, and likely independent of cluster mass and cluster dynamical state. This scenario appears to persist to higher redshifts, but we require a larger and more complete sample to determine this conclusively.
\end{enumerate}

The process by which these BCGs obtain molecular gas is still not clear. We require X-ray observations of our clusters to confirm both their dynamical state and their total mass. Still, as a large, high-z, collection of BCGs selected homogeneously and detected in CO, our sample offers an exciting means by which to study cluster evolution as a whole.

\acknowledgments
TMAW and DAD acknowledge the support of the Natural Sciences and Engineering Research Council of Canada (NSERC) and the Fonds de Recherche Nature et Technologies Quebec (FRQNT).

This paper makes use of the following ALMA data: ADS/JAO.ALMA\#2019.1.01529.S. ALMA is a partnership of ESO (representing its member states), NSF (USA) and NINS (Japan), together with NRC (Canada), MOST and ASIAA (Taiwan), and KASI (Republic of Korea), in cooperation with the Republic of Chile. The Joint ALMA Observatory is operated by ESO, AUI/NRAO and NAOJ. This research has made use of the NASA/IPAC Infrared Science Archive, which is funded by the National Aeronautics and Space Administration and operated by the California Institute of Technology. Based in part on data acquired at the Anglo-Australian Telescope, under program A/2013B/012. We acknowledge the traditional owners of the land on which the AAT stands, the Gamilaraay people, and pay our respects to elders past and present. GW gratefully acknowledges support from the National Science Foundation through grant AST-1517863, from HST program number GO-15294,  and from grant number 80NSSC17K0019 issued through the NASA Astrophysics Data Analysis Program (ADAP). Support for program number GO-15294 was provided by NASA through a grant from the Space Telescope Science Institute, which is operated by the Association of Universities for Research in Astronomy, Incorporated, under NASA contract NAS5-26555.

This project used public archival data from the Dark Energy Survey (DES). Funding for the DES Projects has been provided by the U.S. Department of Energy, the U.S. National Science Foundation, the Ministry of Science and Education of Spain, the Science and Technology Facilities Council of the United Kingdom, the Higher Education Funding Council for England, the National Center for Supercomputing Applications at the University of Illinois at Urbana-Champaign, the Kavli Institute of Cosmological Physics at the University of Chicago, the Center for Cosmology and Astro-Particle Physics at the Ohio State University, the Mitchell Institute for Fundamental Physics and Astronomy at Texas A\&M University, Financiadora de Estudos e Projetos, Funda{\c c}{\~a}o Carlos Chagas Filho de Amparo {\`a} Pesquisa do Estado do Rio de Janeiro, Conselho Nacional de Desenvolvimento Cient{\'i}fico e Tecnol{\'o}gico and the Minist{\'e}rio da Ci{\^e}ncia, Tecnologia e Inova{\c c}{\~a}o, the Deutsche Forschungsgemeinschaft, and the Collaborating Institutions in the Dark Energy Survey.

The Collaborating Institutions are Argonne National Laboratory, the University of California at Santa Cruz, the University of Cambridge, Centro de Investigaciones Energ{\'e}ticas, Medioambientales y Tecnol{\'o}gicas-Madrid, the University of Chicago, University College London, the DES-Brazil Consortium, the University of Edinburgh, the Eidgen{\"o}ssische Technische Hochschule (ETH) Z{\"u}rich,  Fermi National Accelerator Laboratory, the University of Illinois at Urbana-Champaign, the Institut de Ci{\`e}ncies de l'Espai (IEEC/CSIC), the Institut de F{\'i}sica d'Altes Energies, Lawrence Berkeley National Laboratory, the Ludwig-Maximilians Universit{\"a}t M{\"u}nchen and the associated Excellence Cluster Universe, the University of Michigan, the National Optical Astronomy Observatory, the University of Nottingham, The Ohio State University, the OzDES Membership Consortium, the University of Pennsylvania, the University of Portsmouth, SLAC National Accelerator Laboratory, Stanford University, the University of Sussex, and Texas A\&M University.

Based in part on observations at Cerro Tololo Inter-American Observatory, National Optical Astronomy Observatory, which is operated by the Association of Universities for Research in Astronomy (AURA) under a cooperative agreement with the National Science Foundation.

\bibliography{alma_bcgs}{}
\bibliographystyle{aasjournal}

\end{document}